\documentclass[12pt,a4paper]{article}
\usepackage{amssymb,amsmath,graphicx,subcaption,hyperref,cite}
\usepackage{epstopdf}
\usepackage[utf8]{inputenc}
\usepackage{color}
\usepackage[english]{babel}
\begin{document}

\begin{center}
 {\Large\bf Role of Noise in the Fairen-Velarde model of bacterial respiration}
\end{center}
\vskip 1 cm
\begin{center} %
 Soumyadeep Kundu$^{1,2,*}$ and Muktish Acharyya$^{1}$\\
 \textit{$^{1}$Department of Physics, Presidency University, 86/1 College Street, Kolkata-700073, India} 
 \vskip 0.2 cm
 {Email$^1$:sdpknu@gmail.com}
  
 {Email$^2$:muktish.physics@presiuniv.ac.in}
\end{center}
\vspace {1.0 cm}
\vskip 1.5 cm 

\noindent {\large\bf Abstract:} Bacterial respiration, a fundamental biological process, plays a crucial role in ecological systems. The Fairen-Velarde model provides a theoretical framework to study the interplay between oxygen and nutrient concentrations in bacterial populations, representing a system of coupled nonlinear differential equations. In this work, we investigate how the introduction of noise affects the stability and behavior of bacterial respiration. Biological systems are inherently stochastic, with noise arising from environmental fluctuations and molecular-level randomness. Through numerical simulations, we analyze how random fluctuations in oxygen and nutrient concentrations influence the system's stability, particularly the transition between limit cycles and fixed points. Our results demonstrate that noise can induce a reduction in time scales, pushing the system toward a domain of fixed points, which contrasts with the noiseless case where the system exhibits a stable limit cycle. By employing statistical analysis across varying noise intensities, we quantify the likelihood of reaching the fixed domain and examine the area of this domain under different noise conditions. These insights contribute to the broader understanding of how stochastic factors affect bacterial population dynamics, offering implications for microbial ecology and the management of bacterial processes in natural and engineered environments.


\vskip 2cm


\noindent $^*$ Corresponding author\\
\noindent $^{2}$Present Address: \textit{Department of Physics, IIT Bombay, Mumbai-400076, India} \\
\newpage

\section{Introduction}

\noindent Bacterial respiration is a fundamental biological process with profound implications for both ecological dynamics and human health. Bacteria, found in virtually every environment on Earth, play critical roles in nutrient cycling, decomposition, and maintaining the stability and productivity of ecosystems\cite{nut}. For instance, bacteria in the soil contribute to the nitrogen cycle by fixing atmospheric nitrogen, making it available for plant use\cite{nit}, while marine bacteria play a key role in the carbon cycle by decomposing organic matter and facilitating carbon sequestration\cite{carb}. Moreover, bacteria are involved in various symbiotic relationships, providing essential nutrients to their hosts\cite{sym}. However, not all bacterial interactions are beneficial; some bacteria are pathogenic, causing diseases that have significant impacts on human health and agriculture.

One of the fundamental aspects of bacterial respiration is the dynamics of oxygen and nutrient concentrations, which are crucial for bacterial survival, growth, and reproduction\cite{Degn}. Oxygen serves as a terminal electron acceptor in aerobic respiration, a highly efficient energy-producing process, while nutrients provide the necessary substrates for metabolic activities\cite{Camp}. The availability and distribution of these essential resources can significantly influence bacterial population dynamics, including growth rates, community structure, and overall ecosystem function\cite{J}.

In the realm of theoretical biology, mathematical models are invaluable tools for unraveling the complexities of biological systems. These models enable researchers to abstract and simplify real-world processes, providing a framework to systematically analyze the interactions and feedback mechanisms that govern system behavior. The Fairen-Velarde model, proposed in 1979\cite{fairen}, is a prominent example in the field of bacterial respiration. This model provides a mathematical description of the interplay between oxygen and nutrient concentrations in bacterial populations, capturing the essence of their dynamic interactions through a set of coupled nonlinear differential equations. These equations incorporate various parameters, including transport rates, supply rates, and regulatory factors, that shape the behavior and stability of the bacterial population.

In recent years, there has been growing interest in exploring the role of stochastic factors\cite{stot}, such as noise, in biological systems. Biological processes are inherently stochastic, with noise arising from various sources, including environmental fluctuations, random interactions at the molecular level, and demographic variability within populations. Understanding how noise affects the dynamics of bacterial respiration is crucial for gaining insights into the resilience and adaptability of bacterial populations in changing environments\cite{Ocean, Springs, Space}. Noise can result in diverse outcomes, from enhanced robustness and stability to increased variability and unpredictability in population dynamics.

In our previous work, we conducted the analyic calculations for \textcolor{blue}{Fairen}-Velarde model. Additionally, we identified two distinct time-scales in the bacterial respiration process\cite{Kundu-Acharyya}. In this study, we examine the role of noise in the Fairen-Velarde model of bacterial respiration. We focus on understanding how random fluctuations in oxygen and nutrient concentrations affect the stability and behavior of bacterial populations. Through numerical simulations and statistical analysis, we evaluate the effects of strength of the noise and time scale on the dynamics of the system.

Our findings provide insights to a broader understanding of bacterial respiration dynamics and contribute to understand into the mechanisms that underpin the resilience and adaptability of microbial communities. By elucidating the interplay between deterministic and stochastic factors in bacterial respiration, we aim to advance the theoretical foundation of microbial ecology and inform practical strategies for managing bacterial populations in both natural and engineered environments. This integration of mathematical modeling with empirical data may enhance our ability to predict and mitigate the impacts of bacterial activity on ecosystem processes and human health.

In this article, we revise the existing Fairen-Velarde Model\cite{fairen} by introducing noise in section-2. In section-3, we have seen the effects of noise. In section-4, we have analysed those results. The paper ends with a summary and concluding remarks in Section-5.

\section{Fairen-Velarde Mathematical model of Bacterial respiration}

\noindent Theoretically, finding an appropriate model is essential in order to analyze the bacterial life cycles. The time dependency of nutrition and oxygen amounts—two essential components for bacterial survival—must be captured by this model. Such dynamical evolution can be modelled by differential equations involving these two variables (nutrient and oxygen concentrations). Furthermore, a differential equation model of this kind ought to reflect the interdependence of these two fundamental variables.

\textcolor{blue}{Let us denote oxygen by $X$ and nutrient by $Y$. $A$ and $B$ represent the source for oxygen and nutrient respectively.
    \begin{align}    
    {A \underset{k_2}{\stackrel{k_1}{\rightleftharpoons}} X}\\
    {B \xrightarrow {k_3} Y}
    \end{align}
    }
 The most well-known model for bacterial respiration in this context is the Fairen-Velarde (1979) model\cite{fairen}.
\begin{align}
    \frac{dx}{dt} &= k_1 a-k_2 x -k_4 \frac{xy} {1+k_5 x^2} &=f(x,y) \nonumber \\
    \frac{dy}{dt} &= k_3 b -k_4 \frac{xy} {1+k_5 x^2} &=g(x,y).
    \label{fv}
\end{align}

\noindent In this case, the concentration of oxygen in the system is $x(t)$, whereas the concentration of nutrients is $y(t)$. The oxygen transport rate from the chamber to the system is denoted by $k_1$, and the reverse transport rate \textcolor{blue}{of oxygen from the system to the chamber} is represented by $k_2$. The rate at which nutrients are supplied from the chamber to the system is $k_3$. The limit cycle depends on $k_4$ and $k_5$. The concentrations of nutrients and oxygen in the chamber are $a$ and $b$, respectively.

\textcolor{blue}{As a usual practice, to study\cite{strogatz} the linearized version of nonlinear equations, we begin with the nondimensionalisation of the parameters and variables\cite{Kundu-Acharyya} 
\noindent Let us introduce the new set of variables as,
    \begin{equation*}
        \tau=k_2t; ~~~~\hat{x}=\frac{k_4}{k_2}x;~~~~ \hat{y}=\frac{k_4}{k_2}y; ~~~~\alpha=\frac{k_1 k_4}{k_2^2}a;~~~ \beta=\frac{k_3 k_4}{k_2^2}b;~~~~ \kappa=\frac{k_2^2}{k_4^2}k_5        
    \end{equation*}
    Here, the dimensions of $k_1$, $k_2$, $k_3$, $k_4$ and $k_5$ are inversely proportional to time. From our redefined variables, we can infer that the dimension of time $t$, has changed to that of $\tau$ (dimensionless); where as the other variable sets ($\hat{x}$, $x$), ($\hat{y}$, $y$), ($\alpha$, $a$), ($\beta$, $b$) and ($\kappa$, $k_5$) retain their dimensions.
   The redefinition nondimesionalizes the time and 
   the original Fairen-Velarde coupled nonlinear differential equations become:
    }
    \begin{align}
        \frac{d\hat{x}}{d\tau}&=\alpha-\hat{x}-\frac{\hat{x}\hat{y}}{1+\kappa \hat{x}^2}=f_1(\hat{x}, \hat{y}) \nonumber\\
        \frac{d\hat{y}}{d\tau}&=\beta-\frac{\hat{x}\hat{y}}{1+\kappa \hat{x}^2}=f_2(\hat{x}, \hat{y})
    \label{dimless-fv}
    \end{align}
\subsection{Ranges of parameters}
\textcolor{blue}{The fixed point is obtained by setting\\
\begin{align}
    f_1(x,y)&=0\\
    f_2(x,y)&=0
\end{align}
Such that, the coordinates of the fixed point($x^*$, $y^*$) become,
\begin{align}
    x^*&=\alpha-\beta\\
    y^*&=\frac{\beta(1+\kappa(\alpha-\beta)^2}{\alpha-\beta}
\end{align}
The significance of the fixed point and the limit cycle can be explained as follows: a fixed point provides the constant value of oxygen and nutrient concentrations. This implies that the system is arrested in this phase, preventing further bacterial colonization. On the other hand, a limit cycle can be defined as an isolated closed trajectory\cite{strogatz}. In this phase, if all nearby trajectories converge toward the limit cycle, it is referred to as a stable (or attracting) limit cycle. Such a limit cycle exhibits periodic oscillation without external periodic force. In our case, the limit cycle suggests that the consumption rate of oxygen and nutrients is not constant, indicating the periodic colonization (and decolonization) of bacteria.}\\

Now, we evaluate the parameters $\kappa$, $\alpha$, $\beta$. Since they are positive definite (every term depends on rate constants), our conditions become:
\begin{align}
    0&<\beta<\alpha<2\beta\\
    0&<\kappa < \frac{\beta+(\alpha-\beta)+(\alpha-\beta)^2}{(\alpha-\beta)^2(2\beta-\alpha)}
    \label{final-parameters}
\end{align}
To achieve a stable fixed point, the aforementioned relationships must be upheld. To explore the limit cycle, we extend $\kappa$ beyond the specified range. To verify stability (or limit cycle), we evaluate the parameters across various ranges. \textcolor{blue}{It should be noted, however, that this calculation is performed only up to the linearized terms, which fails to fully capture the system’s nonlinearities, leading to inaccuracies in the precise range of stability. We have shown this in Kundu \& Acharyya (2024)\cite{Kundu-Acharyya}. Using the parameters discussed there, we simulated the nonlinear equation (Eq. \ref{dimless-fv})\cite{strogatz} using the 6-th order Runge-Kutta-Fehlberg method\cite{wheatley}.}

Now we are introducing time-dependent random noise to the system. So the equations are transformed to

    \begin{align}
        \frac{d\hat{x}}{d\tau}&=\alpha-\hat{x}-\frac{\hat{x}\hat{y}}{1+\kappa \hat{x}^2}+\eta(t) \nonumber\\
        \frac{d\hat{y}}{d\tau}&=\beta-\frac{\hat{x}\hat{y}}{1+\kappa \hat{x}^2}+\varepsilon(t)
    \label{dimless-fv-noise}
    \end{align}

Here, the white noise is added with a mean value of zero, or $<\eta(t)>=0$. $<\varepsilon(t)>=0$, $<\eta(t)\eta(t')>=k_1\delta(t-t')$, $<\varepsilon(t)\varepsilon(t')>=k_2\delta(t-t')$, $<\eta(t)\varepsilon(t)>=0$. In the next section, we discuss the results of noise ranging in width from $0.4$ to $1.4$ at various time steps. The noise width distribution is uniform around zero, and in all instances, the noise width remains of the same order.
\section{Numerical Results of Fairen-Velarde coupled nonlinear differential equations}

We have solved eqn. (\ref{fv}) using the \textit{6th order Runge-Kutta-Fehlberg} technique to get the instantaneous value of population (both $x(t)$ and $y(t)$) as a function of time($t$).  A time interval of $d\tau = 0.001$ was chosen to ensure minimal error in the 6th-order RKF method.
\vskip 1cm

\subsection{Numerical results without noise}

\begin{figure*}[htpb]
\centering
\includegraphics[width=0.4\columnwidth]{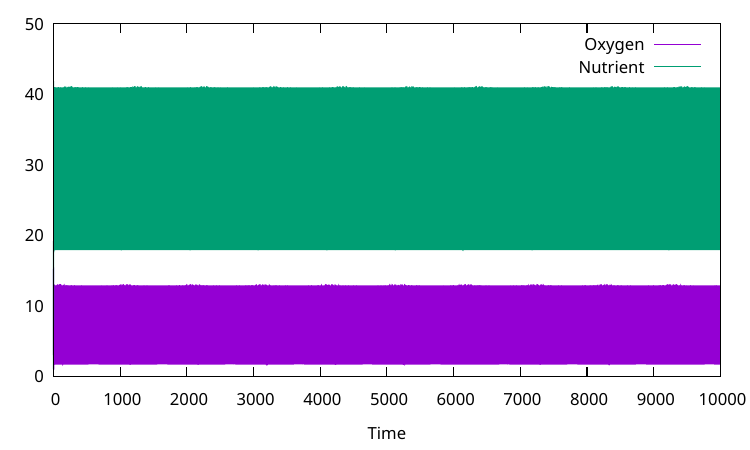}
(a)
\includegraphics[width=0.4\columnwidth]{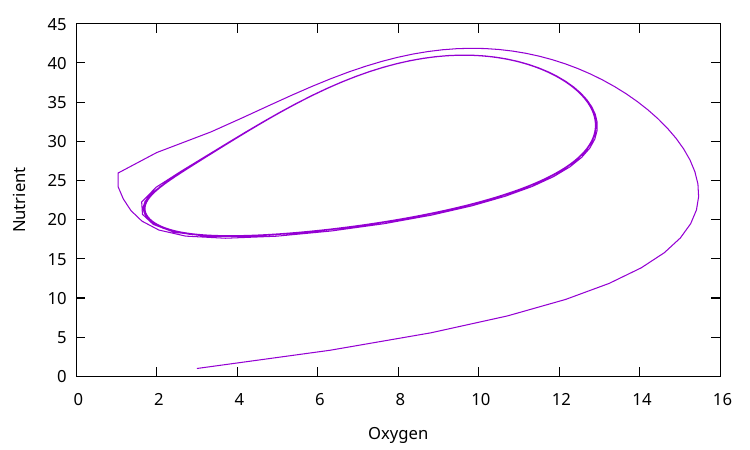}
(b)
\caption{The temporal evolution of the concentrations of the oxygen and nutrients. (a) Concentrations of oxygen and nutrients are plotted as functions of time. (b) The time-eliminated form illustrates the eventual attainment of the limit cycle. Here, $\alpha=19.4$, $\beta=10.997$, $\kappa=0.374$, $\eta=\varepsilon=0$}
\label{fp-simple}
\end{figure*}

We solved Eq. (\ref{dimless-fv-noise}) using the 6th-order Runge-Kutta-Fehlberg method for a fixed set of parameter values. The results are displayed in Fig. \ref{fp-simple}. For this parameter set ($\alpha = 19.4$, $\beta = 10.997$, and $\kappa = 0.374$), we demonstrate the existence of a stable limit cycle in a noiseless system. In this case, we considered a noiseless condition, where $\eta = \varepsilon = 0$.

\vskip 1cm

\subsection{Reduction of Time-Scale in the presence of noise}
By fixing the parameters $\alpha = 19.4$, $\beta = 10.997$, and $\kappa = 0.374$, we created a limit cycle. This limit cycle drives the system toward a fixed point with minimal perturbation in one direction. The system, however, may shift to either side under the influence of random noise with zero mean value. While the noise averages out to zero macroscopically, maintaining equilibrium, we observe that prolonged application of noise with sufficient magnitude can eventually shift the system toward a fixed point. This statistical behavior requires careful analysis using a substantial dataset.

To investigate, we applied random noise and sampled data extensively. For each noise width, we analyzed 1000 ensemble cases. Probabilities were computed for time thresholds $T < 7500$, $T < 10000$, $T < 12500$, $T < 15000$, and $T < 17500$.
\begin{figure*}[h!]
\centering
\includegraphics[width=0.4\columnwidth]{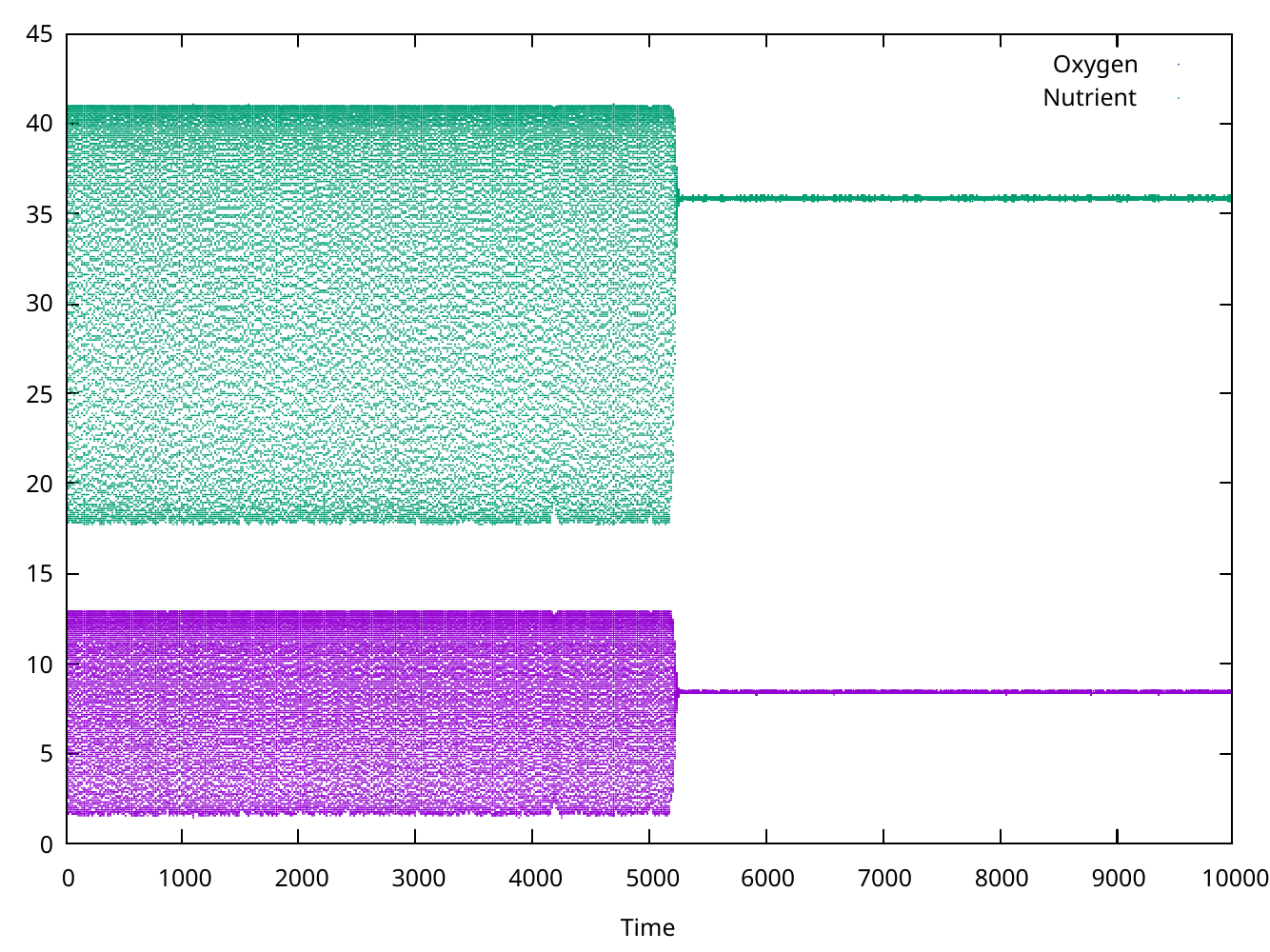}
(a)
\includegraphics[width=0.4\columnwidth]{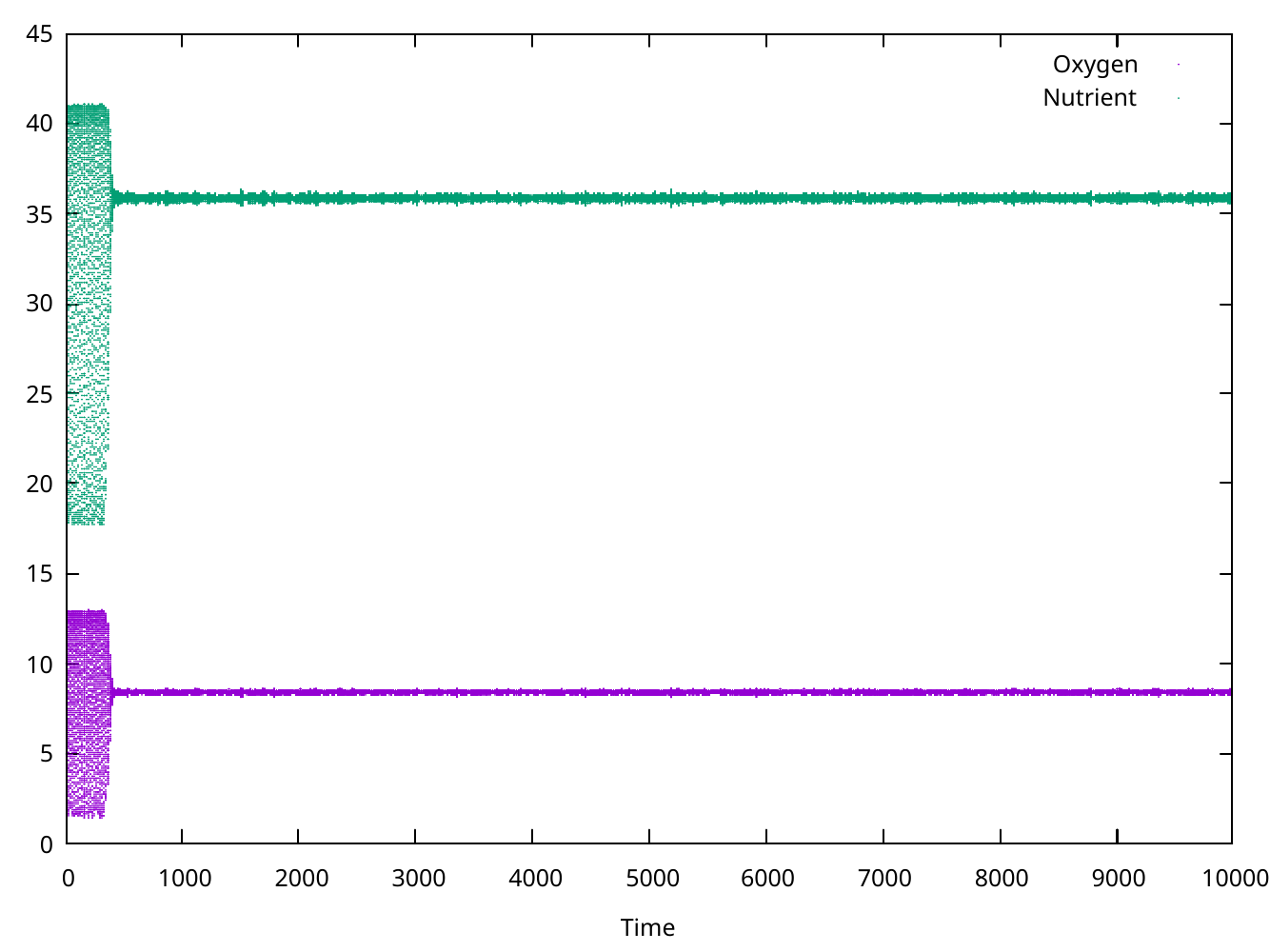}
(b)
\caption{The temporal evolution of the concentrations of the oxygen and nutrients. (a) Noise width is set to 0.4. (b) Noise width is set to 1.4. These figures are for one sample only of each case. The convergence time may be different with different random seeds.}\label{fv-noise-sample}
\end{figure*}

\section{Analysis of the results}

We have talked about the outcomes of our statistical analysis in this part. Reaching a fixed point is not a guarantee, as was previously mentioned. Thus, we have to calculate the likelihood that it has reached the fixed point. Furthermore, Furthermore, we are unable to refer to it as a ``fixed point" because the system is stuck in a fixed region,'' which we have referred to as a ``domain of fixed points" in the material that follows, rather than reaching a ``unique" point. We designate it as a fixed domain if the successive peak-deep value of the nutrient and oxygen is 0.02 times that of the maximum peak-deep value. We have also examined how the domain's area varies with noise width. Lastly, we have investigated the approximate time required to reach the fixed point for each noise width.

\subsection{Probability to reach the domain of fixed points}

We have measured the probability ($P$) of arriving at a given domain in various time frames. It appears that after a cutoff noise, it undoubtedly converges to the domain of fixed points. We are seeing a steady increase in $P$ at various noise widths($w$). Our datapoints have now been fitted using the equation;
 \begin{equation}
 	P(w)=a\rm{tanh}(bw+c)+d
 \end{equation}
 

\begin{figure*}[htpb]
\centering
\includegraphics[width=0.8\columnwidth]{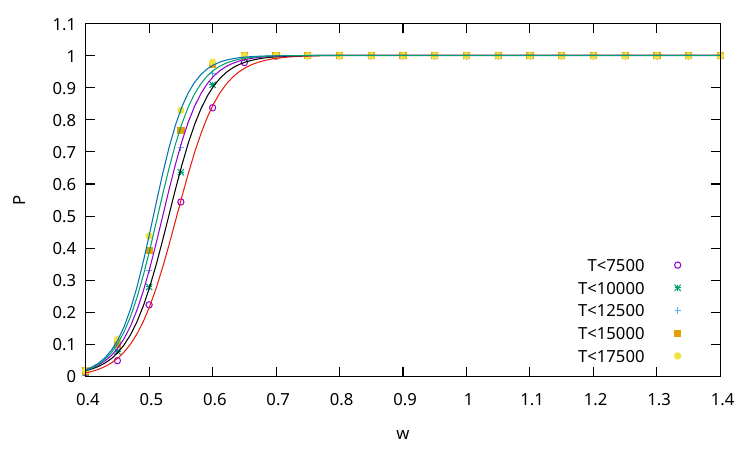}
\caption{Probability of reaching to domain of fixed points is plotted against the noise width. The system is definitely going to the domain of fixed points at higher noise intensity.}
\label{probability}
\end{figure*}

\subsection{Area of the domain of fixed points}
\begin{figure*}[htpb]
\centering
\includegraphics[width=0.8\columnwidth]{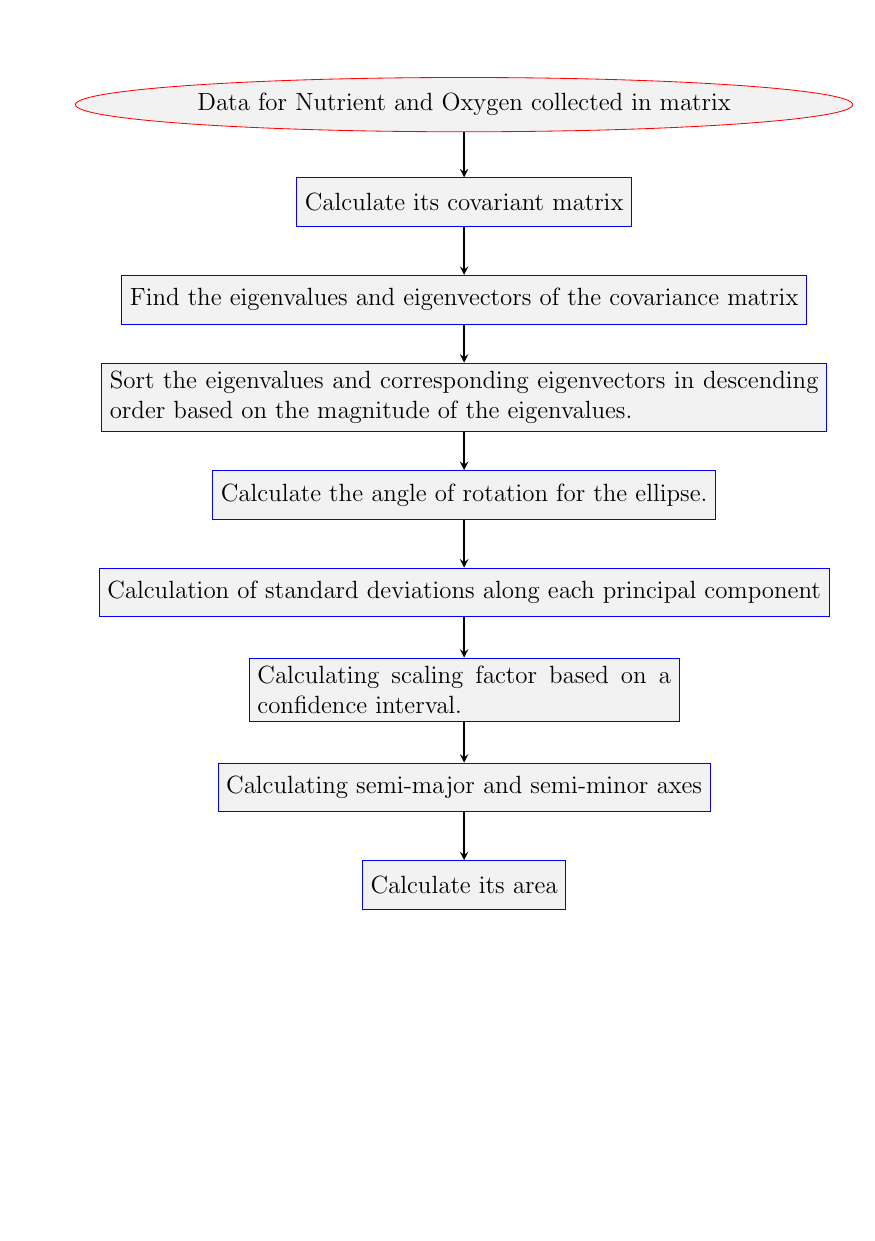}
\caption{Algorithm to determine the area of the ellipse.}
\label{Elipse-alg}
\end{figure*}

The system does not reach a fixed position. Rather, it is creating an elliptical region. We are now exploring the process of determining the area of the region. We employed the Probabilistic Principal Component Analysis approach\cite{Bishop}. We utilized the algorithm described in Fig. \ref{Elipse-alg}\cite{Bishop-book} for this purpose, analyzing data within the following matrix form:

\begin{equation}
X = \begin{bmatrix}
    x_{11} & x_{12} & \ldots & x_{1k} \\
    x_{21} & x_{22} & \ldots & x_{2k} \\
    \vdots & \vdots & \ddots & \vdots \\
    x_{N1} & x_{N2} & \ldots & x_{Nk}
\end{bmatrix}
\end{equation}

We computed the covariance matrix, an essential tool for analyzing relationships between variables in random vectors. The covariance matrix ($C$) is defined as follows:

\begin{equation}
C=\frac{1}{N-1}(X-\bar{X})^T.(X-\bar{X})
\end{equation}
 where $\bar{X}$ is the mean matrix (non-squared in general). This mean is calculated over various realisations of 
 noise of fixed intensity. $\bar{X}$ is defined as;
\begin{equation}
	\bar{X} = \begin{bmatrix}
    \bar{x}_{11} & \bar{x}_{12} & \ldots & \bar{x}_{1k} \\
    \bar{x}_{21} & \bar{x}_{22} & \ldots & \bar{x}_{2k} \\
    \vdots & \vdots & \ddots & \vdots \\
    \bar{x}_{N1} & \bar{x}_{N2} & \ldots & \bar{x}_{Nk}
\end{bmatrix}
\end{equation}
Now, we utilized the standard technique to determine the eigenvalues and eigenvectors of $C$ matrix. Now, if V is the eigenvector and $\lambda$ is the eigenvalue,

\begin{align}
CV=\lambda V \nonumber \\
\Rightarrow  CV=\lambda I V\nonumber \\
\Rightarrow (C-\lambda I)V=0
\end{align}

From this, we can write,
\begin{equation}
det(C-\lambda I)=0
\end{equation}
So, we identified the eigenvalues and subsequently find the eigenvectors from this process. Then we sorted the eigenvalues and corresponding eigenvectors in descending order based on the magnitude of the eigenvalues. \textcolor{blue}{The data points are distributed like an ellipsoid. Now, the eigenvectors determine the axes of the N-dimensional ellipsoid. Such an ellipsoid can be described as,
\begin{equation}
	\sum_{i=1}^N \frac{x_i}{a_i} = 1
	\label{ellipsoid}
\end{equation} 
where, $a_i$ are the corresponding axes of the ellipsoid. It can be determined by the eigenvectors of the covariance matrix. As, it is a $N\times N$ square matrix, we get $N$ eigenvectors in $N$-dimensional space\cite{Ian}. Each eigenvector represents the direction of one principal component. Also, respective eigenvalues correspond to variances along these axes. The eigenvector with the largest eigenvalue corresponds to the first principal component, which is the direction of maximum variance\cite{Gilbert}. Now, in N-dimension, if an eigenvector looks like\\
\begin{equation}
  \bf{V}=[v_1,~v_2,~\cdots,~v_N] 
\end{equation}
the angle with the respective coordinate look like,\\
\begin{equation}
  \theta_i=cos^{-1}\big(\frac{v_i}{||\bf{V}||}\big)
  \label{gen_th}
\end{equation}
Besides, we can also calculate the pairwise angle between eigenvectors by,
\begin{equation}
  cos(\theta_{ij})=\frac{\bf v_i\cdot\bf v_j}{||\bf v_i||~||\bf v_j||}
\end{equation}
So, we have to calculate $\frac{N(N-1)}{2}$ independent angles to describe the ellipsoid. Now, we can also describe that $\Lambda_i$ are eigenvalues of the covariant matrix which also gives the variance of data. So, we can calculate the standard deviation of data by;
\begin{equation}
\sigma_i=\sqrt{\Lambda_i}
\end{equation}
 Now, we can calculate $a_i$ of eq:\ref{ellipsoid} from our standard deviation\cite{Richard}, \\
\begin{equation}
    a_i=r\sigma_i
\end{equation}
where r is a scaling factor that depends on the confidence interval($P_i$). The confidence interval can be calculated by\cite{conf},
\begin{align}
  P_1(r)=erf\Big(\frac{r}{2}\Big)&\rm{~~~~~~for~1D~case~}; i=1  \label{P1}\\
  P_2(r)=1-exp\Big(-\frac{r^2}{2}\Big)&\rm{~~~~~for~2D~case~}; i=2 \label{P2}\\
  P_i(r)=P_{i-2}(r)-\Big(\frac{r}{\sqrt{2}}\Big)^{i-2}\frac{exp(-\frac{r^2}{2})}{\Gamma(\frac{n}{2})}&\rm{~~~~~for~a~general~dimension~} \label{P3}; i>2
\end{align}
We can calculate the volume of N-dimensional ellipsoid by\cite{Ellip},
\begin{equation}
    V=\frac{2}{n}\frac{\pi^{n/2}}{\Gamma(n/2)}\prod_{i=1}^{N}a_i
\end{equation}
}
Now, for our case, we have two variables i.e. Nutrient and Oxygen; denoting as N and O; the matrix look like.
\begin{equation}
 X = \begin{bmatrix}
    N_{1} & O_{1} \\
    N_{2} & O_{2} \\
    \vdots & \vdots \\
    N_{N} & O_{N}
\end{bmatrix}
\end{equation}

Now, the $X$ is a $n\times 2$ matrix, in general a non-square matrix. The covariant matrix $C=X^{T} X$ is a $2\times 2$ square matrix. \textcolor{blue}{So, our covariant matrix looks like,\\
 {\begin{equation} C =\begin{bmatrix}
	Var(N) & Cov(N,O)\\
	Cov(O, N) & Var(O)\\
\end{bmatrix}
\label{Cov}
\end{equation}
}
}
  So, we have two eigenvalues of the covariant matrix $C$, namely, 
 $\lambda_1$ and $\lambda_2$, with two eigenvectors $V_1$ and $V_2$. So the rotation matrix $R$ can be given by eigenvectors as:

\begin{equation} R = \begin{bmatrix}
    V_{1} & V_{2}
\end{bmatrix} =\begin{bmatrix}
	v_{11} & v_{12}\\
	v_{21} & v_{22}\\
\end{bmatrix}
\label{Rot-1}
\end{equation}

Again, \textcolor{blue}{we have to consider that the principal axes of the ellipse may not align with the original co-ordinate axes because the nutrient and oxygen concentration levels are correlated. So, we have to calculate the rotation angle of the semi-major axis($\theta$) with the principal axes.} $R$ can be also written in the terms of rotation angle($\theta$) as,
\begin{equation}
R = \begin{bmatrix}
    cos(\theta) & -sin(\theta) \\
    sin(\theta) & cos(\theta)
\end{bmatrix}
\label{Rot-2}
\end{equation} 
 Now, to get the semi-major axis of the ellipse, we have considered the eigenvector corresponding to the larger eigenvalue, where as the smaller eigenvalue corresponds to the semi-minor axis. We can see that, the two rotation matrices mentioned in Eq. \ref{Rot-1} and \ref{Rot-2} are equivalent. So taking the first column, we can calculate,
 
 \begin{align}
 tan\theta&=\frac{v_{21}}{v_{11}} \nonumber \\
 \Rightarrow \theta&=tan^{-1} \frac{v_{21}}{v_{11}} =cos^{-1} \frac{v_{11}}{\sqrt{v_{11}^2+v_{21}^2}}
 \label{ori}
 \end{align}

Furthermore, from the equations we can say that, if C=0, we can say that the data are uncorrelated and the eigenvectors lie with the original axes, that is, $\theta=0$.

Now we calculated the standard deviation of the principal components to find the dispersion of the set of values. The eigenvalues of the covariance matrix are the \textcolor{blue}{variance} of the principle components; so if we take its square root, we have got the standard deviation. So, mathematically,

\begin{align}
\sigma_x=\sqrt{\lambda_1} \\
\sigma_y=\sqrt{\lambda_2}
\end{align} 

Now, we measured the scaling factor. For this, we took the $99\%$ confidence limit, i.e. we took the $1\%$ data as outliers. We got scaling factor($S$) and confidence limit($C_L$) is related as per eq:\ref{P2},

\begin{align}
S=\sqrt{-2 log (1-C_L)}
\end{align} 

Now, we had $\sigma_x$, $\sigma_y$ and $S$. So using this, we found the semi-major($a$) and semi-minor($b$) axes. The mathematical equation to find it is,
\begin{align}
a=S\sigma_x\\
b=S\sigma_y
\end{align}
So, the area of the desired ellipse($E_A$) calculated as,
\begin{equation}
E_A=\pi a b
\end{equation}

\begin{figure*}[h!]
\centering

\includegraphics[width=0.29\columnwidth]{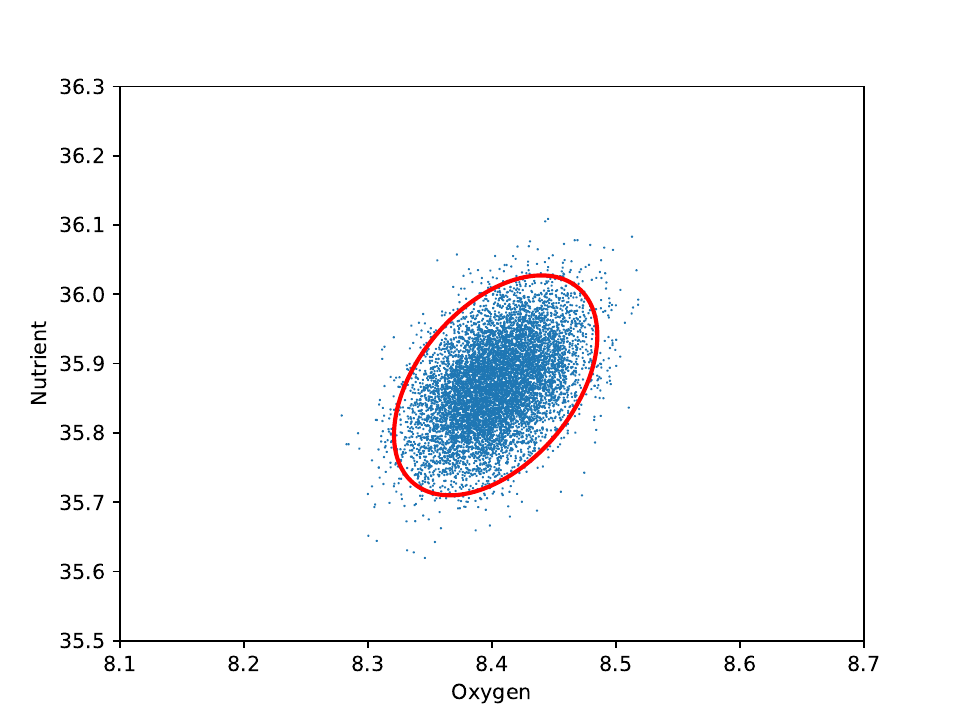}
(a)
\includegraphics[width=0.29\columnwidth]{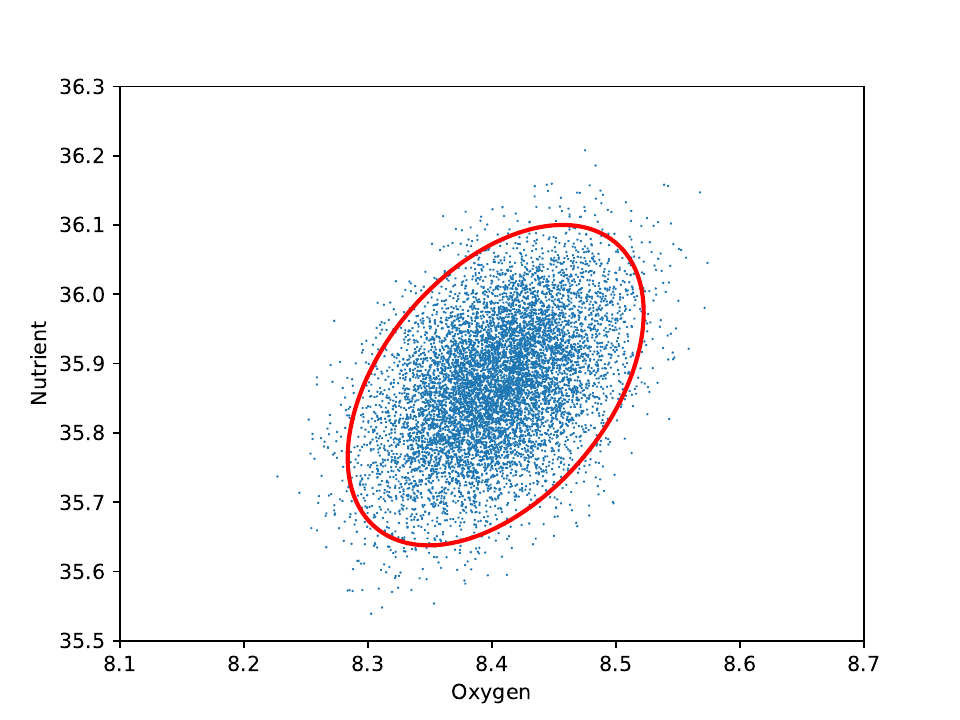}
(b)
\includegraphics[width=0.29\columnwidth]{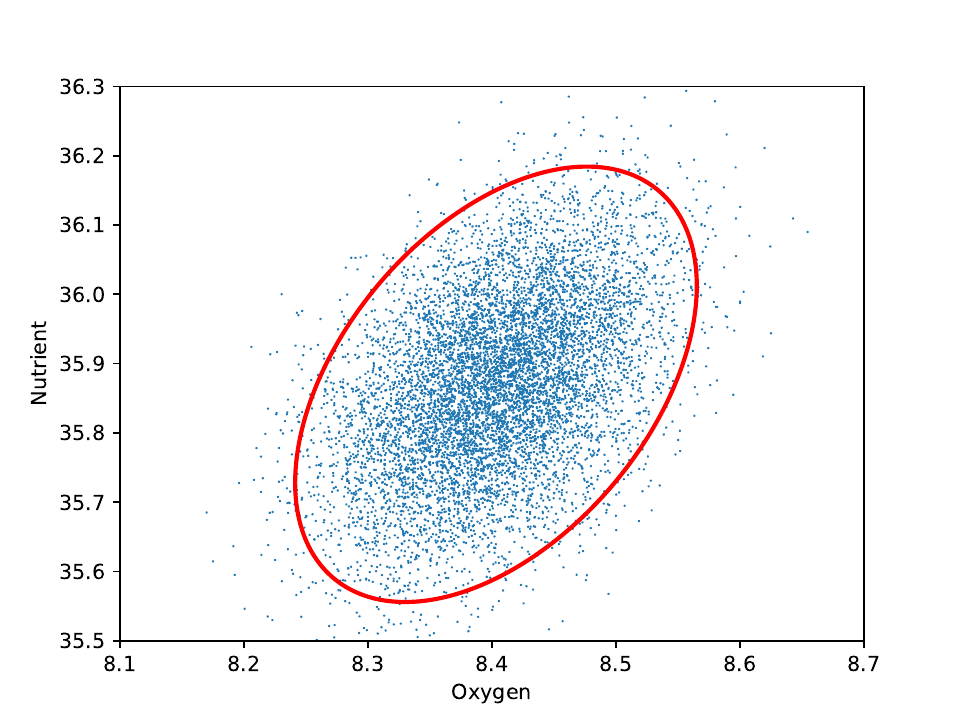}
(c)
\includegraphics[width=0.29\columnwidth]{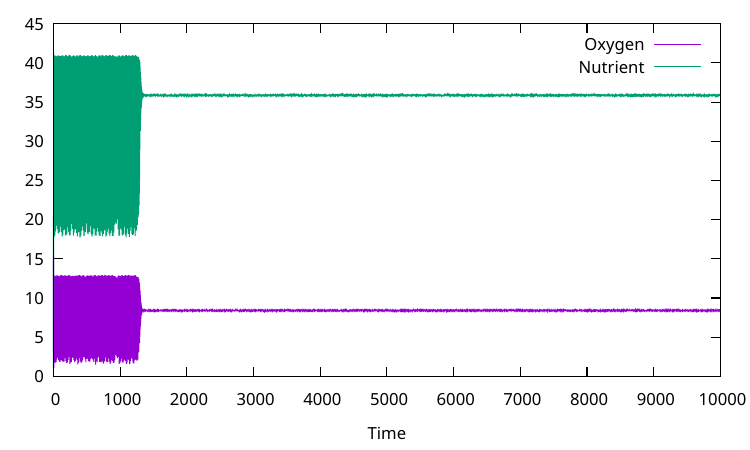}
(d)
\includegraphics[width=0.29\columnwidth]{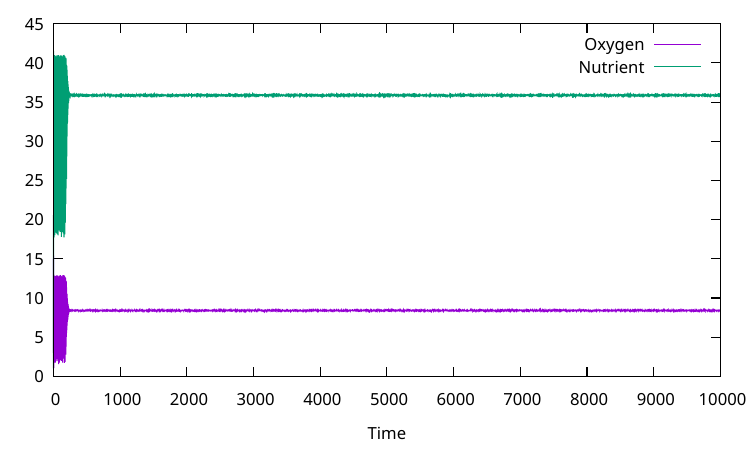}
(e)
\includegraphics[width=0.29\columnwidth]{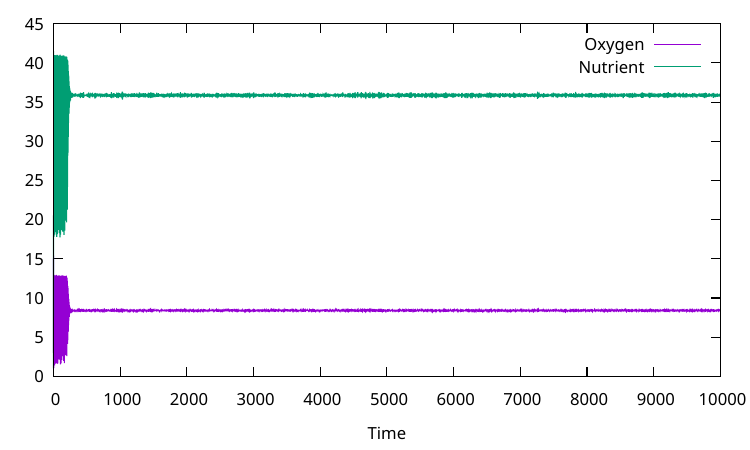}
(f)

\caption{(a-c)The evolution of size of elliptical region with increase in noise. (a) Noise width is set to $0.8$ (b) Noise width is set to $1$. (c) Noise width is set to $1.4$.\textcolor{blue}{(d-f)The temporal evolution of the concentrations of the oxygen and nutrients. (d) Noise width is set to $0.8$ (e) Noise width is set to $1$. (f) Noise width is set to $1.4$. These figures are for one sample only of each case. The convergence time may be different with different random seeds.}}\label{Elipse-size}
\end{figure*}
When we increase the noise, we observed that the area gradually increases. For increased noise, we observed an asymptotic convergence of the data points to a straight line. Thus, we used the equation \ref{area-fit} to fit the data points for noises larger than $1$.

\begin{equation}
E_A(w)=mw+c \rm{, where~} m=1.11; c=-0.65
\label{area-fit}
\end{equation} 

\begin{figure*}[htpb]
\centering
\includegraphics[width=0.8\columnwidth]{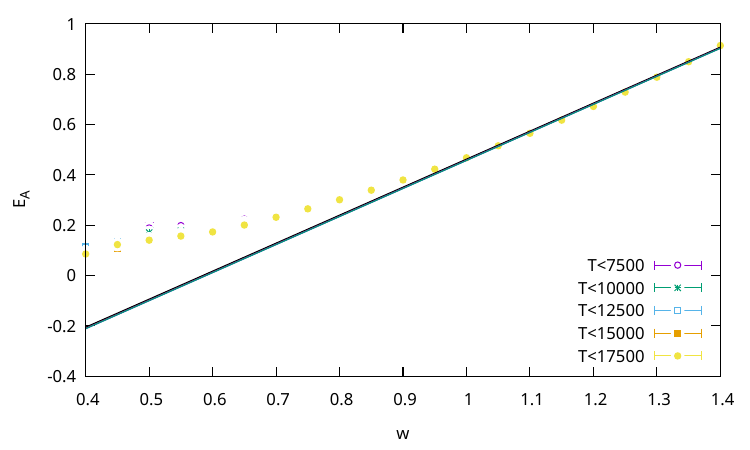}
\caption{Area of fixed domain is plotted with noise intensity. The area is asymptotically increasing with increase in noise intensity.}
\label{area-noise}
\end{figure*}

\subsection{Time to reach the domain of fixed points}
For each noise level, we determined the average time ($t_{avg}$) required to reach the fixed domain within specified constraints. For averaging, only ensembles reaching the defined domain were included. A peak in time is observed between noise levels $0.45$ and $0.5$. To model this behavior more precisely, we fit it to the equation below:
\begin{equation}
t_{avg}(w) = a e^{-\frac{(w - b)^2}{2c^2}}+d.
\label{Time=fit}
\end{equation}

\begin{figure*}[htpb]
\centering
\includegraphics[width=0.8\columnwidth]{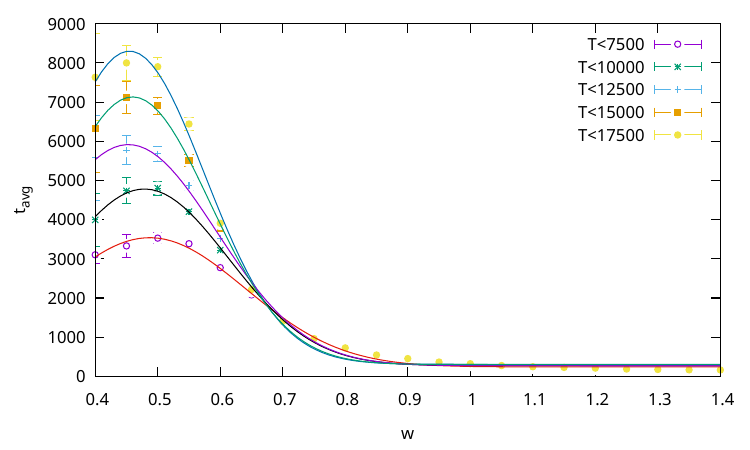}
\caption{Average time of reaching to domain of fixed points is plotted with noise intensity.}
\label{time}
\end{figure*}

We have observed that the system is obviously going to the fixed domain for higher noises asymptotically. There is a saturation in timescale at higher noises. It is taking time around 250-300 to reach the fixed domain.

\newpage

\section{Summary and concluding remarks}

The Fairen-Velarde model is a prototypical model of bacterial respiration. The two-dimensional deterministic nonlinear differential equation for the oxygen and the nutrient contents can 
grasp the essential features of bacterial respirations. Depending on the various domains of the parameters, it can 
successfully provide 
the active phase (limit cycle) and the inactive phase (fixed point) of bacterial respiration. The existence of 
two time scales before reaching the fixed point has been shown in our previous work published in Int. J. Mod. Phys C, 35 (2024) 2450094.

In this article, we primarily focus on the role of noise in the steady-state behavior of bacterial respiration. Stochastic effects, introduced by noise in the modified Fairén-Velarde model, lead to several interesting phenomena. We observe that noise accelerates the system's motion toward the fixed point, effectively reducing the gross time scale required to reach it. This finding has significant implications in bacteriology, as introducing noise—whether thermal, impurity-based, or mechanical—can expedite the transition of a bacterial system into the inactive phase (fixed point) compared to the noiseless case.

However, the presence of noise alters the nature of the fixed point. Instead of a single fixed point, the scattering of data creates a domain of fixed points. The size of this domain increases with the intensity of the noise, growing asymptotically in a linear manner. Additionally, in the presence of noise, reaching the domain of fixed points becomes probabilistic over a given time frame. The probability of reaching this domain within a specified time exhibits a hyperbolic tangent-like behavior as a function of noise intensity.

\vskip 0.5cm

\noindent {\bf Acknowledgements:}

\noindent Soumyadeep Kundu thankfully acknowledges Md. Samsul Habib Shah, IIT Madras, for having valuable discussions about the principal component analysis method. \textcolor{blue}{We thank the anonymous reviewer for important suggestions.}

\noindent {\bf Authors' contribution:}

\vskip 0.2cm

\noindent \textit{Soumyadeep Kundu} has developed the code for numerical calculations, collected data, prepared figures and written the manuscript. \textit{Muktish Acharyya} has conceptualized the problem, analysed the results and written the manuscript. 

\vskip 0.3cm

\noindent {\bf Data availability statement:}
The data will be available on reasonable request to Soumyadeep Kundu.

\vskip 0.3cm

\noindent {\bf Conflict of interest statement:} We declare that this manuscript is free from any conflict of
interest. The authors have no financial or proprietary interests in any material discussed in this article.

\vskip 0.3cm
\noindent {\bf Funding statement:} No funding was received particularly to support this work.

\newpage 

\newpage

\end{document}